\documentclass[twocolumn,showpacs,preprintnumbers,amsmath,amssymb,amsfonts,final,a4paper,aps, citeautoscript,footinbib,pra,superscriptaddress]{revtex4-1}
\usepackage[pdftex]{graphicx,color}
\usepackage{dcolumn}
\usepackage{bm}
\usepackage[latin1]{inputenc}
\usepackage{mathrsfs}
\usepackage{epstopdf}
\usepackage{times}
\usepackage{float}
\usepackage[sort&compress]{natbib}
\usepackage[colorlinks,bookmarks=true,citecolor=blue,linkcolor=blue,urlcolor=blue, breaklinks=true]{hyperref}
\begin{document}
\title{Bulk effective theory in coupled wire construction from generalized Wilson line}
\author{Yukihisa Imamura}
 \email{yukihisa.imamura@yukawa.kyoto-u.ac.jp}
\affiliation{Division of Physics and Astronomy, Graduate School of Science, Kyoto University, 
Kyoto 606-8502, Japan.}
\affiliation{%
 Yukawa Institute for Theoretical Physics, Kyoto University, Kyoto 606-8502, Japan}%
\author{Keisuke Totsuka}%
\affiliation{%
 Yukawa Institute for Theoretical Physics, Kyoto University, Kyoto 606-8502, Japan}%
\begin{abstract}
We reconsider the coupled wire construction, which is a useful method of obtaining two and three-dimensional 
topologically ordered states from an array of one-dimensional CFTs, 
and show that there is the hidden structure in the bulk of the constructed model.  
In order to uncover the structure hidden in the gapped bulk,  
we introduce a generalized Wilson line and show the necessity of introducing a new gauge field, 
which behaves like the Chern-Simons gauge field. 
From these examination, it is possible to show that in this formulation the bulk theory is the U(1) Chern-Simons gauge theory and the edge theory is the chiral Luttinger theory.
We explicitly demonstrate our method for the Laughlin states and the chiral spin liquid state. 
\end{abstract}
\maketitle
\section{\label{sec:level1}Introduction}
Better understanding of topological order \cite{WenText} is one of the most fundamental and important 
problems in modern condensed matter physics.
The Abelian fractional quantum Hall state, which is now called the Laughlin state, was discovered experimentally 
in 80's \cite{TsuiStormerGossard} and since then many other topologically ordered states have been 
proposed theoretically
\cite{Moore1991,Read1999,KalmeyerLaughlin1987,Wen1989,Wen1991PRB44,MoessnerSondhi,Kitaev2003,LevinWen}. 
Since these states are featureless in the bulk,  
it is not possible to distinguish them by local order parameters.
Therefore, we inevitably use an intrinsically non-local description.
It is necessary to formulate an inherently non-local and non-perturbative field theory in order to gain a deeper understanding of topological order.

In this paper, we focus on an approach to the construction of topologically ordered states 
initiated by Kane and his collaborators \citep{KaneMukhopadhyayLubensky}, which is called the coupled-wire construction (CWC).
Specifically, it is a method of obtaining two-dimensional topologically ordered states 
from an array of one-dimensional wires put on a plane by introducing interactions among them. 
It is possible to construct various topologically ordered states by changing the one dimensional theories 
and tailoring the interactions to the desired states.  
Recently, CWC has been employed to construct various two- and three-dimensional topological states.   
The list includes the Abelian and non-Abelian fractional quantum Hall states \cite{KaneMukhopadhyayLubensky,TeoKane} 
and the chiral spin liquid states \cite{Meng2015}.  
We can also use CWC to construct topological insulators and superconductors \citep{Qi2011}
in two- and three dimensions in the similar way \cite{Sagi2015,Santos2015,Sagi2014,Seroussi2014}.  
In particular, the derivation of the periodic table of integer and fractional fermionic topological phases and the entanglement structure analysis are discussed \cite{Neupert2014}.

However, there are two problems in the conventional formulation of CWC: 
first of all, the local gauge invariance is broken after introducing the interactions that are necessary 
to impose constraints among the wires and obtain non-trivial topological orders.
This is not quite satisfactory and we have to improve it.
Second point is that the original formulation \cite{KaneMukhopadhyayLubensky, TeoKane} focuses only on 
the edge effective theory and how the bulk effective theory emerges is not very clear. 
Namely, the fermion or boson fields are defined on the individual wires, 
whereas there is no fields in the space among wires. 
This problem is equivalent to the lack of the (two-dimensional) bulk theories and is intrinsically due to the formalism.

In this paper, we show that the two problems mentioned above are resolved by introducing the Peierls phase and considering the Wilson line \cite{Wilson1974} between wires. 
For this purpose, it is convenient to adopt the Lagrangian formulation that enables us to discuss symmetries easily.
The combination of the Wilson line and the fermions is a gauge invariant object 
and is often used in nuclear physics for the description of the quark confinement.
Note that we need to generalize the conventional form of the Wilson line 
as it is not quite convenient for our formulation of CWC.   
With the generalized form of the Wilson line, a bulk theory naturally appears in the space among the wires.

What is necessary for us to conclude that the constructed state is in fact in some topological ordered state?
First, we need to identify the fundamental degrees of freedom in the construction, e.g., charge, spin, etc.
Second, the identification between the effective theory constructed and the conventional effective theory.
One of the most critical features of the effective theories of topologically ordered states is 
the bulk/edge correspondence \cite{Witten1989,Wen1990ChiralLuttinger}.  
In our construction, the effective theories in the bulk and in the edge are derived simultaniously, 
so we may identify the constructed state with the topologically ordered state more persuasively than the original formulation.

This paper is organized as follows.
In Sec. II, we will try to reformulate CWC by taking the Laughlin states for an example.
The role of the Wilson line will be explained and both the bulk effective theory and the edge one will be derived.
Moreover we will discuss the excitation and the quasiparticle in the state, and derive the filling factor 
hierarchical structure.
In Sec. III, another example, the chiral spin liquid state, is obtained in the similar way of the Laughlin state.

\section{Laughlin state in Coupled Wire Constructon}
\label{sec:CWC-for-Laughlin}
The Laughlin state has been a paradigmatic example in the study of topological order.
It has been experimentally realized in two-dimensional electron systems 
in a strong magnetic field \cite{TsuiStormerGossard}.
Theoretically, we can describe the basic properties of this state with 
the effective field theories and the bulk-edge correspondence.
For the Laughlin state with the filling factor $\nu=1/m$, 
the effective theory in the bulk is the level-$m$ U(1) Chern-Simons gauge theory \cite{Blok1990,Lopez1991}, 
and that in the edge is 
the chiral Luttinger-liquid theory \cite{Wen1991ModernPhysicsLettrsB}.

The construction of the Laughlin state by using CWC has been proposed by Kane et al. \cite{KaneMukhopadhyayLubensky}
and the method has been extended later to the non-Abelian fractional quantum Hall cases 
by Teo and Kane \cite{TeoKane} and other states with topological order.
However, the formulation is not gauge-invariant and 
lacks the viewpoint of the bulk effective Chern-Simons theory.
In order to understand the emergence of the non-trivial topological order in the two-dimensional {\em bulk} 
from the stacking of one-dimensional wires,
we recast it into a gauge-invariant formulation.
In this section, we will explain the reformulated CWC for the Laughlin state 
and show how the effective theories in the bulk and the edge appear in this construction.
It will be also shown that the gapped excitation only exists in the bulk and it is Abelian anyons.

\subsection{From single wire to coupled wires}
\begin{figure}
\includegraphics[width=80mm]{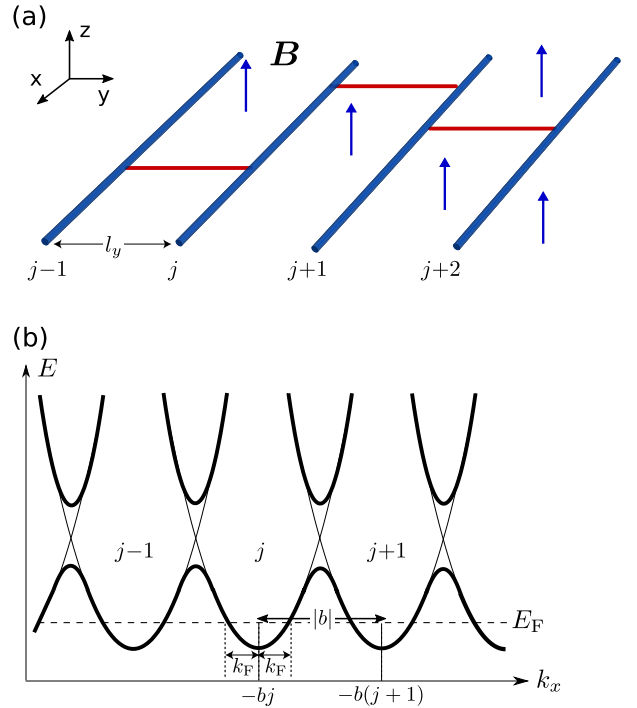} 
\caption{(Color online) %
(a)
An array of coupled wires in a perpendicular magnetic field.
The wires interact with their nearest neighbors thorough the Wilson lines, which are shown with the red lines.
(b)
The energy band structure of the coupled wires.
A perpendicular magnetic field in the Landau gauge shifts 
the quadratic dispersion relations of the individual wires (labelled by $j$) in a $j$-dependent manner: $k \to k + bj$. 
The interactions between the adjacent wires cause the gaps at the crossing points 
and the Landau-level structure is formed. 
If we set the Fermi energy $E_\text{F}$ below the gap (as shown by dashed line), 
the Fermi momentum $k_\text{F}$ and the filling factor $\nu$ can be defined well. 
\label{fig:wires}}
\end{figure}
Let us begin with the one-dimensional theory defined on each wire.
The one-dimensional wires running in the $x$-direction 
are aligned in parallel on the $xy$-plane with the inter-wire distance $a$, as is shown in Fig.~\ref{fig:wires}(a). 
A strong magnetic field $B$ applied perpendicular to the plane is incorporated by 
the gauge potential $\mathbf{A}=(-By,0)$ (Landau gauge)
\footnote{
We do not need to fix it to the Landau gauge.
The momentum shift is in general not uniform and the order of the dispersion relations in the momentum space is not equal to that of the wires in the real space.
In this case, we need to put the Wilson lines by the order in the momentum space.
}. 
We consider a non-relativistic spinless fermion (with mass $M$) moving on each wire (labelled by $j$) 
with the dispersion relation $E_{j}(k_x) = (k_x + bj)^2 / 2M$ with $b=eaB/\hbar$ [see Fig.~\ref{fig:wires}(b); 
$e<0$ for electrons and hence $b<0$].   
The inclusion of interactions among the neighboring wires opens the gaps at the crossing points, 
which are identified with the gaps separating the Landau levels.
At the filling fraction $\nu <1$ where the lowest Landau level is partially filled, the Fermi momentum 
$k_\text{F}$ is well-defined and related to $\nu$ as $\nu=2k_\text{F} / |b|$.   
After the linearization of the low-energy dispersion around the two Fermi points, we have two Dirac fermions 
$R_j$ (around $k_x= - bj+k_{\text{F}}$) and $L_j$ ($k_x= - bj- k_{\text{F}}$)  
that are related to the original spineless fermion $c_{j}(x)$ as \footnote{%
In contrast to the usual bosonization treatment, we have included the rapidly oscillating parts 
$e^{i(-bj \pm k_{\text{F}}) x}$ 
in the definition of $R_j$ and $L_j$.} 
\begin{equation}
c_{j}(x) \approx R_{j}(x) + L_{j}(x)  \; .  
\label{eqn:spinless-F-by-R-L}
\end{equation}
The low-energy effective action for the $j$-th wire reads as 
\begin{equation}
S_{0}^{(j)} = \int d^2x \, v^{0}_{\text{F}} \bar{\psi_j} i \gamma^{\mu} \left( \partial_{\mu} 
- i \frac{e}{\hbar} \mathcal{A}^{(j)}_{\mu}
- i \mathcal{K}_{\text{F},\mu} \gamma_5\right) \psi_j  \; ,
\label{eqn:Dirac-action}
\end{equation}
where the summation over the repeated Greek indices $\mu=0,1$ is implied and $d^{2}x=dtdx$.  
The velocity given by $v^{0}_{\text{F}} =\hbar k_{\text{F}}/M$ is common to all wires.  
The Dirac fermions $\psi_j$ and $\bar{\psi}_j$ have the following representation,
\begin{equation}
\psi_j =
\begin{pmatrix}
R_j \\
L_j
\end{pmatrix}
, \qquad \bar{\psi}_j = \psi^{\dagger}_j \gamma^0 .
\end{equation}
and we have introduced the following Dirac gamma matrices:
\begin{equation}
\begin{split}
& \gamma^0 =
\begin{pmatrix}
0 & 1 \\
1 & 0
\end{pmatrix}
,~~~~~ \gamma^1 =
\begin{pmatrix}
0 & -1 \\
1 & 0
\end{pmatrix} 
\\
& 
\gamma_5 =\gamma^0 \gamma^1 =
\begin{pmatrix}
1 & 0 \\
0 & -1
\end{pmatrix}. 
\end{split}
\end{equation}
The Dirac fermions are minimally coupled to the U(1) gauge field 
$\mathcal{A}^{(j)}_\mu = (0, A^{(j)}_x) = (0, -Baj)$ induced 
by the perpendicular magnetic field and 
$\mathcal{K}_{\text{F},\mu} = (0, - k_\text{F})$ has been introduced to take care of the $R/L$-dependent shift 
from the reference momentum $k^{(j)}_{x}= - bj$. 

Using the Abelian bosonization,  
a spinless fermion on each chain ($j$) can be expressed, in low energies, in terms of the bosons 
$\phi_{j,\text{L/R}}$ as  
\begin{equation}
\begin{split}
& R_{j} = \frac{\kappa_{j}}{\sqrt{2\pi a_0}} e^{i(-bj +k_{\text{F}})x}
\exp\left(i  \phi_{j,\text{R}}\right)  
\\
& L_{j} = \frac{\kappa_{j}}{\sqrt{2\pi a_0}} e^{i(-bj - k_{\text{F}})x}
\exp\left(i \phi_{j,\text{L}}\right)  \; ,
\end{split}
\label{eqn:spinless-boson2fermion}
\end{equation}
where $\kappa_{j}$ are the Klein factors necessary for the anti-commutation among the spineless 
fermions on different wires.  
Then, in terms of the bosonic fields $\Phi_{j}$ (compactified on a circle with radius $1$) 
and their duals $\Theta_{j}$
\begin{equation}
\Phi_{j}=\frac{1}{2}(\phi_{j,\text{R}}+\phi_{j,\text{L}}) \; , \quad 
\Theta_{j}= \frac{1}{2}(\phi_{j,\text{R}}-\phi_{j,\text{L}}) \; ,
\label{eqn:PhiTheta2chiral}
\end{equation}
the low-energy effective Hamiltonian for the spineless fermion on the $j$-th chain is given by
\begin{equation}
{\cal H}^{(j)}_{\text{Dirac}} = \frac{v^{0}_{\text{F}}}{2\pi}\int\!dx \sum_{j} 
\left\{
\frac{1}{K_j}: \!(\partial_{x}\Phi_{j})^{2}\! :
+K_j : \! (\partial_{x}\Theta_{j})^{2}\! :
\right\}  \; .
\end{equation}
The Luttinger-liquid parameter $K_j$ equals to 1 for the non-interacting case and may be 
modified in the presence of interactions. 

We next would like to introduce the interactions among the wires.
Let us consider only a pair of wires $j$ and $(j+1)$.
In gapping out most of the gapless degrees of freedom and leaving only the gapless 
chiral Luttinger liquids at the edges, the authors of 
Refs.~\cite{KaneMukhopadhyayLubensky,TeoKane} introduced the following elaborate inter-wire interaction 
allowed by U(1) and translation symmetries:
\begin{equation}
\mathcal{H}_{\text{int}}
= - g \sum_{j}
: \left( L_{j+1}^{\dagger}  R_j \right) ^{\frac{m+1}{2}} 
\left( L_j^{\dagger} R_{j+1} \right)^{\frac{m-1}{2}} :
\label{eqn:inter-wire-int-by-fermion}
\end{equation}
where $m$ is an odd integer.  
Here, we are not very precise about the ordering of the fermion operators; 
we assume that they are correctly ordered in such a way that, when bosonized, they reproduce \eqref{eqn:inter-wire-int}. 
Roughly, this interaction is made up of simultaneous $\frac{m-1}{2}$-pled (intra-wire) backscattering 
on the adjacent wires and inter-wire single-particle hopping (i.e., $\mathcal{H}_{\text{int}}$ consists of 
$m$-particle processes). 
On the other hand, in terms of the bosonic fields, $\mathcal{H}_{\text{int}}$ reads as 
\begin{equation}
\begin{split}
& \mathcal{H}_{\text{int}} 
= - g \sum_{j} \int dx : \cos \left[ \Phi_{j+1} - \Phi_{j} - m (\Theta_{j+1} + \Theta_{j}) \right] : \\
& = - g \sum_{j} \int dx : \cos \left[ (\Phi_{j+1} - m \Theta_{j+1}) - (\Phi_{j} + m \Theta_{j} ) \right] :  \; .
\end{split}
\label{eqn:inter-wire-int}
\end{equation}
Physically, the part $\Phi_{j+1} - \Phi_{j}$ comes from the single-particle hopping between neighboring wires, 
while $m (\Theta_{j+1} + \Theta_{j})$ from the tailored backscattering {\em within} individual chains 
\cite{KaneMukhopadhyayLubensky,TeoKane}. 

However, this interaction is not gauge invariant and 
we have to modify it by introducing the Peierls phase \cite{Santos2015}:
\begin{equation}
\begin{split}
&\mathcal{H}_{\text{int}}
= - g \sum_{j}
\left( L_{j+1}^{\dagger} e^{i \int dy \frac{e}{\hbar} A_y } R_j \right)^{\frac{m+1}{2}} \\
&~~~~~~~~~~~~~~~~~~~~~~~~~~~~~~~
\times \left( L_{j}^{\dagger} e^{i \int dy (- \frac{e}{\hbar} A_y )} R_{j+1} \right)^{\frac{m-1}{2}} + \text{h.c.}  \\
& = - g \sum_{j}
\left( L_{j+1}^{\dagger} R_{j+1} \right)^{\frac{m-1}{2}}
\left( L_{j}^{\dagger} R_{j} \right)^{\frac{m-1}{2}} \\
& ~~~~~~~~~~~~~~~~~~~~~~~~~~~~~~~~~~~~~~~
\times \left[ L_{j+1}^{\dagger} e^{i \int dy \frac{e}{\hbar}  A_y} R_j \right] + \text{h.c.}  \; ,
\end{split}
\label{eqn:Peierls-phase}
\end{equation}
where $A_y$ is the $y$-component of the gauge field. 
This modification is equivalent to that in Ref.~\cite{Santos2015}.  

In the strong-coupling limit ($g \gg 1$), the fields are no longer independent but are pinned  
to obey the local constraints
\begin{equation}
\left(\Phi_{j+1}(x) - \Phi_{j}(x)\right) - m \left(\Theta_{j+1} (x)+ \Theta_{j}(x)\right) = \int^{j+1}_j dy \frac{e}{\hbar}  A_y  \; ,
\label{eqn:pinning}
\end{equation}
which is the key to the coupled-wire construction of the Laughlin state. 
Note that here the ``local'' means the boson fields $\Phi$ and $\Theta$ depend on the coordinate $x$ in the each wire. 

\subsection{Generalized Wilson line}
\label{sec:stability}

In gauge theories, especially in QCD, the Wilson line is used to construct a gauge-invariant non-local correlator 
between hadrons (see, e.g., \citep{Wilson1974}).   
For instance, the combination
\begin{equation}
\bar{\psi}(x_1)  \exp  \left[ i \int^{x_1}_{x_0} dx^\mu A_\mu \right]\psi (x_0) 
\label{eqn:standard-Wilson-line}
\end{equation}
is obviously invariant under the usual local gauge transformation
\begin{equation}
\psi \rightarrow e^{i\varphi(x)} \psi ,~~ \bar{\psi} \rightarrow \bar{\psi} e^{-i\varphi(x)},
~~A_\mu \rightarrow A_\mu +  \partial_\mu \varphi .
\end{equation}
We would like to use the Wilson line in order to examine the relation between the wires and derive the feature of the two-dimensional bulk theory, which might be the Chern-Simons gauge theory.

However, instead of using Eq.~\eqref{eqn:standard-Wilson-line},  
we adopt a slightly different form of the Wilson line:
\begin{equation}
\bar{\psi}_{j+1} \exp \left[ i\int^{(j+1)a}_{ja} dy 
\left( \frac{e}{\hbar} A_y + a_y \gamma_5  \right) \right]\psi_j,
\label{eqn:modified-Wilson-line}
\end{equation}
where $y$ is the coordinate specifying the position {\em between} the wires $j$ and $j+1$.  
The $y$-component of the gauge field $A_y$ already appeared in Eq.~\eqref{eqn:Peierls-phase}.  
On the other hand, the $a_y$, which, for the moment, is just an unknown field, 
will turn out to be the U(1) Chern-Simons gauge field in Sec.~\ref{sec:bulk-FT-Laughlin}. 

Having introduced the Dirac fermions $\psi_j$ and $\bar{\psi}_{j+1}$ describing the individual wires, 
let us consider the symmetries under which the entire system is kept invariant. 
Before introducing the interaction \eqref{eqn:inter-wire-int-by-fermion}, only forward scattering interactions 
exist and the system is invariant under a diagonal U(1) transformation and a global chiral U(1) transformation: 
\begin{equation}
\psi_j \rightarrow e^{i\varphi_j} \psi_j ,~~ \bar{\psi}_j \rightarrow \bar{\psi}_j e^{-i\varphi_j} 
\quad \text{[diagonal U(1)]} 
\label{eqn:diagonal-U1}
\end{equation}
and
\begin{equation}
\psi_j \rightarrow e^{i\theta_j \gamma_5} \psi_j ,~~~~~ \bar{\psi}_j \rightarrow \bar{\psi}_j e^{i\theta_j \gamma_5} 
\quad \text{[chiral U(1)]}
\; ,
\label{eqn:chiral-U1}
\end{equation}
respectively.   
From the expressions \eqref{eqn:spinless-boson2fermion} and \eqref{eqn:PhiTheta2chiral}, 
it is clear that the two U(1) operations \eqref{eqn:diagonal-U1} and \eqref{eqn:chiral-U1} 
respectively correspond to the translation for the fields $\Phi_j$ and $\Theta_j$:
\begin{equation}
\Phi_j (x) \to \Phi_j (x) + \varphi_j \;\; \text{and} \; \; 
\Theta_j (x) \to \Theta_j (x) + \theta_j  \; .
\label{eqn:gauge-bosonized}
\end{equation}
For simplicity, here we only consider the global (i.e., $x$-independent) gauge transformation 
on each wire, while keeping the $j$-dependence of $\{ \varphi_{j},\theta_{j} \}$.
We can readily generalize it to the fully local gauge transformation 
by introducing the gauge field $A_x$. 
The $y$-component $A_{y}$ is less trivial as the system is discrete in the stacking ($y$) direction.  
In this case, as has been discussed in the previous section, one may use the Peierls phase or the Wilson line 
to introduce $A_{y}$.

Now let us consider the fate of the gauge transformations \eqref{eqn:gauge-bosonized} 
in the strong-coupling limit. 
In doing so, we first note that, in this limit, the above two gauge transformations must satisfy 
the following constraints in order for the pinning condition \eqref{eqn:pinning} to be kept invariant:
\begin{equation}
\theta_{j+1} + \theta_{j} = \frac{1}{m} \left(\varphi_{j+1} - \varphi_{j} \right)  .
\label{eqn:U1-constraint}
\end{equation}
In other words, the essence of the inter-wire interaction 
is to lock the independent parameters $\{ \varphi_{j},\theta_{j} \}$ 
by imposing the constraint \eqref{eqn:U1-constraint} between the two gauge transformations \eqref{eqn:diagonal-U1} 
(diagonal) and \eqref{eqn:chiral-U1} (chiral) thereby leaving only a subset of the gauge transformations.  
In the following, we think of this residual symmetry as responsible for some emergent gauge structure 
(in the $y$-direction). 

We begin with the modified inter-wire interaction \eqref{eqn:Peierls-phase}. 
As the chiral transformation \eqref{eqn:chiral-U1} is constrained by the relation \eqref{eqn:U1-constraint},
it transforms as follows:
\begin{equation}
\begin{split}
&\mathcal{H}_{\text{int}}
= - g \sum_{j}
\left( L_{j+1}^{\dagger} e^{i \int dy  \frac{e}{\hbar} A_y  } R_j \right)^{\frac{m+1}{2}} \\
&~~~~~~~~~~~~~~~~~~~~~~~~~~~~~~~
\times \left( L_{j}^{\dagger} e^{i \int dy (- \frac{e}{\hbar} A_y )} R_{j+1} \right)^{\frac{m-1}{2}} + \text{h.c.}  \\
& \xrightarrow{\eqref{eqn:chiral-U1}} - g \sum_{j}
\left( L_{j+1}^{\dagger} e^{i \int dy \frac{e}{\hbar} A_y + i(\theta_j + \theta_{j+1})} R_j \right)^{\frac{m+1}{2}} \\
&~~~~~~~~~~~~~~~~~~
\times \left( L_{j}^{\dagger} e^{i \int dy (- \frac{e}{\hbar} A_y ) + i(\theta_j + \theta_{j+1})} R_{j+1} \right)^{\frac{m-1}{2}} + \text{h.c.}  \\
& = - g \sum_{j}
\left( L_{j+1}^{\dagger} R_{j+1} \right)^{\frac{m-1}{2}}
\left( L_{j}^{\dagger} R_{j} \right)^{\frac{m-1}{2}} \\
& ~~~~~~~~~~~~~~~~~~~
\times \left[ L_{j+1}^{\dagger} e^{i \int dy \frac{e}{\hbar} A_y + i(\varphi_{j+1} - \varphi_j)} R_j \right] + \text{h.c.} \\
& = - g \sum_{j}
\left( L_{j+1}^{\dagger} R_{j+1} \right)^{\frac{m-1}{2}}
\left( L_{j}^{\dagger} R_{j} \right)^{\frac{m-1}{2}} \\
& ~~~~~~~~~~~~~~~~~~~
\times \left[ L_{j+1}^{\dagger} e^{i \int dy \frac{e}{\hbar} (A_y +\frac{\hbar}{e} \partial_y \varphi)} R_j \right] + \text{h.c.} 
\; ,
\end{split}
\end{equation}
where, in passing to the final result, we have used the constraint \eqref{eqn:U1-constraint} to trade 
$\theta_j + \theta_{j+1}$ with $\varphi_{j+1} - \varphi_j$.  
We can see here that the chiral invariance of the interaction \eqref{eqn:inter-wire-int-by-fermion}  
is replaced with the usual gauge transformation 
\begin{equation}
A_y \rightarrow A_y + \frac{\hbar}{e} \partial_y \varphi 
\label{eqn:A-transformation}
\end{equation}
through the constraint \eqref{eqn:U1-constraint}.   
Therefore, one sees that, as far as the interaction \eqref{eqn:inter-wire-int-by-fermion} is 
concerned, the chiral transformation 
is no longer independent and may be expressed as the ordinary vector transformation on $A_y$.  

However, this is not the case for other (Wilson-line-like) operators and this necessitates us 
introducing another gauge field which governs the low-energy physics even without the external electromagnetic 
field $A_\mu$. 
To see this, let us consider the usual Wilson line:
\begin{equation}
\bar{\psi}_{j+1} \exp \left[ i\int^{(j+1)a}_{ja} dy 
\left(  \frac{e}{\hbar} A_y  \right) \right]\psi_j.
\label{eqn:Wilson-line}
\end{equation}
The constraint condition \eqref{eqn:U1-constraint} connects the gauge transformation \eqref{eqn:diagonal-U1} 
and the chiral transformation \eqref{eqn:chiral-U1} for the fermion fields $\psi$ and $\bar{\psi}$. 
On the other hand, 
if we consider the chiral transformation \eqref{eqn:chiral-U1} on the usual Wilson line \eqref{eqn:Wilson-line}, 
we have to consider the $\gamma_5$ term in the exponential.  
This suggests us to consider the following {\em generalized} Wilson line: 
\begin{equation}
\bar{\psi}_{j+1} \exp \left[ i\int^{(j+1)a}_{ja} dy 
\left( \frac{e}{\hbar} A_y + a_y \gamma_5  \right) \right]\psi_j \; .
\label{eqn:def-modified-Wilson}
\end{equation}

In general, this is not invariant under the chiral transformation either. Nevertheless, {\em with the constraint}  
\eqref{eqn:U1-constraint}, we can make it chiral-invariant  
\begin{equation}
\begin{split}
&\bar{\psi}_{j+1} \exp \left[ i\int^{(j+1)a}_{ja} dy \left( \frac{e}{\hbar}A_y + a_y \gamma_5 \right) \right]\psi_j 
\xrightarrow{\eqref{eqn:chiral-U1}}
\\
 & 
\bar{\psi}_{j+1} \exp \left[ i\int^{(j+1)a}_{ja} dy \left(
 \frac{e}{\hbar} A_y + a_y \gamma_5 \right) + i \left( \theta_{j+1} + \theta_j \right) \gamma_5  \right] \psi_j  \\
& = \bar{\psi}_{j+1} \exp \Biggl[ i\int^{(j+1)a}_{ja} \! \! \! \! \! \! \! dy \left(  \frac{e}{\hbar} A_y  +  a_y \gamma_5 \right) \\
& \phantom{ \bar{\psi}_{j+1} \exp \Biggl[ i\int^{(j+1)a}_{ja} \! \! \! \! \! \! \! dy  \frac{e}{\hbar} A_y } + i \frac{1}{m} \left( \varphi_{j+1}  - \varphi_j \right) \gamma_5  \Biggr] \psi_j 
\\
& = \bar{\psi}_{j+1} \exp \left\{ i\int^{(j+1)a}_{ja} dy \left[ 
 \frac{e}{\hbar} A_y +  \left( a_y 
+ \frac{1}{m}  \partial_y \varphi \right) \gamma_5 \right]  \right\} \psi_j \; ,
\end{split}
\end{equation} 
provided that $a_{y}$ transforms properly under the chiral transformation \eqref{eqn:chiral-U1}:
\begin{equation}
a_{y} \to a_y 
+ \frac{1}{m}  \partial_y \varphi   \; .
\label{eqn:a-transformation}
\end{equation}
We call the operator defined in Eq.~\eqref{eqn:def-modified-Wilson} the generalized Wilson line.

Note that just to keep the Hamiltonian invariant, we do not need to introduce the gauge field $a_y$.
In fact, we first note that 
the generalized Wilson line \eqref{eqn:def-modified-Wilson} is written in terms of $L$ and $R$ as
\begin{equation}
\begin{split}
L^{\dagger}_{j+1} \exp i\int dy & \left( \frac{e}{\hbar} A_y + a_y \right) R_j \\
&+ R^{\dagger}_{j+1} \exp i\int dy \left( \frac{e}{\hbar} A_y - a_y \right) L_{j} \; .
\end{split}
\end{equation}
Then, we are led to adopting, instead of \eqref{eqn:Peierls-phase}, the following:
\begin{equation}
\begin{split}
&\mathcal{H}_{\text{int}}
= - g \sum_{j}
\left( L_{j+1}^{\dagger} e^{i \int dy ( \frac{e}{\hbar} A_y +a_y)  } R_j \right)^{\frac{m+1}{2}} \\
&~~~~~~~~~~~~~~~~~~~~~~~~~~~~~~~
\times \left( L_{j}^{\dagger} e^{i \int dy (- \frac{e}{\hbar} A_y + a_y)} R_{j+1} \right)^{\frac{m-1}{2}} + \text{h.c.}  \\
& = - g \sum_{j}
\left( L_{j+1}^{\dagger} R_{j+1} \right)^{\frac{m-1}{2}}
\left( L_{j}^{\dagger} R_{j} \right)^{\frac{m-1}{2}} \\
& ~~~~~~~~~~~~~~~~~~~
\times \left[ L_{j+1}^{\dagger} e^{i \int dy ( \frac{e}{\hbar} A_y + ma_y)} R_j \right] + \text{h.c.} \; .
\end{split}
\label{eqn:Peierls-phase-2}
\end{equation}  
Therefore, as long as the constraint \eqref{eqn:U1-constraint} is satisfied, 
the transformations \eqref{eqn:A-transformation} and \eqref{eqn:a-transformation} {\em equally} do the job 
in order to keep the Hamiltonian \eqref{eqn:Peierls-phase-2} invariant under \eqref{eqn:chiral-U1}.  

%
\subsection{Bulk effective theory}
\label{sec:bulk-FT-Laughlin}
What we have to do next is to examine the physical role of the gauge field $a_y$ in this formalism.
The Dirac fermions \eqref{eqn:Dirac-action} defined on the individual wires are subject to 
the ``external fields'' $\mathcal{A}^j_\mu$ and $\mathcal{K}^{\text{F}}_\mu$.
These fields can be formally eliminated from the covariant derivative by the following mixed transformation
\begin{equation}
\begin{split}
& \psi_j \rightarrow e^{i\int dx^\mu \left( \frac{e}{\hbar} \mathcal{A}^{(j)}_\mu 
+ \mathcal{K}^{\text{F}}_\mu \gamma_5 \right)} \chi_j , \\
& \bar{\psi}_j \rightarrow \bar{\chi}_j e^{i\int dx^\mu 
\left(- \frac{e}{\hbar} \mathcal{A}^{(j)}_\mu + \mathcal{K}^{\text{F}}_\mu \gamma_5 \right)}
\end{split}
\end{equation}
or, equivalently, by 
\begin{equation}
R_{j} \to e^{i (-bj + k_{\text{F}})x} R_{j} \; , \;\;
L_{j} \to e^{i (-bj - k_{\text{F}})x} L_{j} 
\label{eqn:gauge-tr-Laughlin}
\end{equation}
as
\begin{equation}
S_{0}^{(j)} \rightarrow \int d^2x \bar{\chi}_{j} i \gamma^{\mu} \partial_{\mu} \chi_{j}  .
\end{equation}
This transformation also changes the form of the operator \eqref{eqn:modified-Wilson-line} as
\begin{equation}
\begin{split}
&\bar{\psi}_{j+1} \exp \left[ i\int^{(j+1)a}_{ja} dy 
\left( \frac{e}{\hbar} A_y +  a_y \gamma_5  \right) \right]\psi_j
\\
& \rightarrow \bar{\chi}_{j+1}
\exp \biggl\{ i\int^{(j+1)a}_{ja} dy \int dx \\
& \qquad 
\left[ \frac{e}{\hbar} B + 2\pi \left(\frac{1}{2\pi} \partial_x a_y + \frac{k_\text{F}}{\pi a} \right) \gamma_5 \right]
 \biggr\}  \chi_j .
\end{split}
\label{eqn:Wilson-line-transformed}
\end{equation}
Since the integral $\int dy dx$ is taken over the two-dimensional region between the two wires 
$j$ and $(j+1)$, all the functions that are to be integrated are defined in the two-dimensional bulk.
The quantity $k_\text{F}/(\pi a)$
may be viewed as the area-density 
of the fermion and therefore we expect that $\partial_x a_y / (2\pi)$ plays 
the role of the density fluctuation.  
As the density is the temporal component of the conserved current in $(2+1)$-dimensional space-time, 
which is parametrized as $J^{\mu} \propto \epsilon^{\mu\nu\rho}\partial_{\nu}a_{\rho}$ \cite{Frohlich-Z-91}, 
the correct form of the density should read as: 
\begin{equation}
\rho = \frac{\kappa}{2\pi} \epsilon^{ij} \partial_{i}a_{j} = \frac{\kappa}{2\pi}(\partial_{x}a_{y} -\partial_{y}a_{x}) \; ,
\end{equation}
with an unknown constant of proportionality $\kappa$.

From this interpretation, we can derive the Chern-Simons term as follows.
In the next subsection, following Refs.~\cite{KaneMukhopadhyayLubensky, TeoKane}, 
we will define the bulk quasiparticles by the $2\pi$ shift of the field $\Phi$ (and hence $\varphi$).
This definition and the constraint \eqref{eqn:U1-constraint} show that the electric charge of the quasiparticle   
is fractional: $e/m$. 
Therefore we may write the relation between the fluctuating part of the electron density $\rho$
and the gauge field $a_y$ as
\begin{equation}
\frac{1}{2\pi} \epsilon^{ij} \partial_{i} a_{j} = \frac{1}{m} \rho  \; .
\label{eqn:Chern-Simons1}
\end{equation}
Taking time-derivative of both sides of \eqref{eqn:Chern-Simons1} 
and using the continuity equation of the fermion current, 
$\partial_0 \rho + \partial_{i}  J^{i} = 0$, we obtain
\begin{equation}
\partial_{i} \left\{ 
\frac{1}{2\pi} \epsilon^{ij} \partial_{0}  a_{j} + \frac{1}{m} J^{i} 
\right\}  = 0  \; , 
\label{eqn:Chern-Simons2}
\end{equation}
which then implies 
\begin{equation}
\frac{1}{2\pi} \epsilon^{ij} \partial_{0}  a_{j} + \frac{1}{m} J^{i} = \epsilon^{ij} \partial_{j} f 
\end{equation}
for a certain function $f$.  
If we define $a_{0} \equiv 2\pi f$, we can write the current as
\begin{equation}
J^{\mu} = \frac{m}{2\pi} \epsilon^{\mu\nu\rho}\partial_{\nu}a_{\rho}  \; .
\label{eqn:CS-EoM}
\end{equation}
The Lagrangian that yields the relations \eqref{eqn:CS-EoM} 
as the equations of motion is known to be 
the level-$m$ U(1) Chern-Simons gauge theory \cite{Dunne-LesHouches-98}:
\begin{equation}
\mathcal{L}_{\text{CS}} = \frac{m}{4\pi} \epsilon^{\mu \nu \rho} a_{\mu} \partial_{\nu} a_{\rho} - a_{\mu} J^{\mu}.
\label{eqn:Chern-Simons3}
\end{equation} 
Therefore, we may conclude that $a_y$ appearing in the Wilson line \eqref{eqn:def-modified-Wilson} 
is the $y$-component of the U(1) Chern-Simons gauge field.  
Recently, the effective Chern-Simons theory \eqref{eqn:Chern-Simons3} has been derived 
in the context of coupled-wire approach to fractional Chern insulators by calculating the electromagnetic 
response \cite{Santos2015}.  Here we derived the same result by considering the residual symmetry 
that survives the inter-wire coupling and the current conservation.  

Moreover, we can extract the information of the filling factor from the form of the Wilson line 
\eqref{eqn:Wilson-line-transformed}. 
The ratio of $eB$ to $2k_\text{F}/a$ have to $1/m$
by the effect of the constraint between a gauge and a chiral transformation. 
Therefore the filling factor is
\begin{equation}
\nu = \frac{2k_\text{F}}{b} = \frac{1}{m}.
\end{equation}
\subsection{Quasi-particle excitations}
\begin{figure}
\includegraphics[width=80mm]{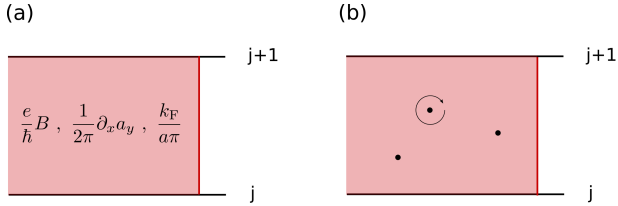} 
\caption{(Color online) 
(a) The existence of the Chern-Simons gauge theory in the two-dimensional region 
between a pair of adjacent wires.  The filling factor is tuned by the interaction.  
The magnetic field, the rotation of the Chern-Simons gauge field and the area-density of the fermion are defined in the two-dimensional bulk.
(b) Quasiparticles in the bulk space.
The moving quasiparticle around the others is affected the phase change through the background Chern-Simons gauge field, so the quasiparticles obey fractional statistics.
\label{fig:CS-quasiparticle}}
\end{figure}

Finally, let us discuss the excitations and the filling factor hierarchy. 
We have so far considered an arbitrary pair of adjacent wires and the Wilson line between them.
However, there are several wires on a plane, so we should include the effect of all of them.
In the case where the number of the wires is finite (e.g. $N_y$), 
the first and the last wires correspond to the edge and the others to the bulk.

All low-energy fluctuations are suppressed in the bulk by the suitably designed interactions \eqref{eqn:inter-wire-int} 
and we can show \cite{KaneMukhopadhyayLubensky,TeoKane} 
that the only low-energy excitations exist near the edges and are described by the chiral-Luttinger-liquid theories 
\cite{Wen-CLL-90,Wen-edge-91,Wen-92}. 
The key in the derivation of the bulk Chern-Simons theory was the relation \eqref{eqn:U1-constraint} that 
constrains possible shift of the fields.  
However, due to the periodicity of the cosine-interaction \eqref{eqn:inter-wire-int}, 
we may allow for the following shift for possible (strong-coupling) ground states:  
\begin{equation}
\begin{split}
&\Theta_{j+1} + \Theta_{j} = \frac{1}{m} \left(\Phi_{j+1} - \Phi_{j} \right)\\
& \rightarrow \Theta_{j+1} + \Theta_{j} = \frac{1}{m} \left\{ \Phi_{j+1} - \Phi_{j} + 2\pi Q_{j,j+1}  \right\} \; ,
\end{split}
\end{equation}
where $Q_{j,j+1}$ is an arbitrary integer. 
From this, one may suspect that there are infinitely many degenerate ground states corresponding 
to all the possible shifts of the above form.  
However, careful analysis similar to that used in Refs.~\cite{Lin-B-F-98,Lecheminant-T-06-SU4} 
shows that, when the system is periodic 
in the $y$-direction (i.e., 2-torus), any shifts satisfying $\sum_{j=1}^{N_y} Q_{j,j+1}=0$ (mod $m$; $N_y$ being 
the number of wires) can be absorbed into 
the gauge redundancy
\begin{equation}
\phi_{j,\text{L}} \to \phi_{j,\text{L}} + 2\pi N_{j,\text{L}} \; , \;\; 
\phi_{j,\text{R}} \to \phi_{j,\text{R}} + 2\pi N_{j,\text{L}} \quad 
(N_{j,\text{L,R}} \in \mathbb{Z})  
\end{equation}
of the fermions [see Eq.~\eqref{eqn:PhiTheta2chiral}] and do not generate physically distinct states.  
This implies that there exist precisely $m$ gauge-inequivalent ground states as is predicted by 
the direct analysis of the Abelian Chern-Simons gauge theories on a torus \cite{Wen-89,Wen-N-90}.  

We can consider the gapped excitations in the bulk as well. 
In order to create the quasi-particle excitations, let us consider the following solitonic configuration \cite{TeoKane}:
\begin{align}
&\Theta_{j+1} + \Theta_{j} = \frac{1}{m} \left(\Phi_{j+1} - \Phi_{j} \right)\\
& \rightarrow \Theta_{j+1} + \Theta_{j} = \frac{1}{m} \left\{ \Phi_{j+1} - \Phi_{j} 
+ 2\pi q_{\text{QP}} \vartheta_{\text{step}}(x-x_q^{(j)})  \right\} \; ,
\end{align}
where $\vartheta_{\text{step}}(x-x_q^{(j)})$ denotes the step function. 
In order to see that this configuration in fact corresponds to a quasiparticle between the two wires, 
let us consider again the Wilson line \eqref{eqn:def-modified-Wilson}.  
By repeating the same steps as in Sec.~\ref{sec:bulk-FT-Laughlin}, 
we see that the additional term $2\pi q_{\text{QP}} \vartheta_{\text{step}}(x-x_q^{(j)})$ shifts the density term in 
Eq.~\eqref{eqn:Wilson-line-transformed} as
\begin{equation}
\frac{1}{2\pi} \partial_x a_y 
\to 
\frac{1}{2\pi} \partial_x a_y + \frac{1}{m} q_{\text{QP}} \delta(x -x_q^{(j)}) \; , 
\label{eqn:soliton}
\end{equation}
where the delta functions comes from the derivative of the step function.  
By the argument in Sec.~\ref{sec:bulk-FT-Laughlin}, 
this implies that the solitonic configuration \eqref{eqn:soliton} creates a charge $q_{\text{QP}}/m$ 
fractionalized particle between the two wires $j$ and $j+1$. 
The fractional statistics of the quasiparticles  can be understood from the fractional charge. 
If a quasiparticle move through a closed loop $C$ around others, it feels the Aharonov-Bohm phase of the magnetic field and the quasi-particle fluxes $\frac{1}{m} q_{\text{QP}} \delta(x -x_q^{(j)})$ in the region enclosed in the loop $C$.
The latter effect causes the phase shift $2\pi N/m$, where $N$ is the number of the quasi-particles in the region.
In the case of the exchange of two quasi-particles, the phase is $\pi /m$ and this shows the mutual fructional statistics of quasi-particles.

It is possible to construct the hierarchical states \cite{Haldane1983,Halperin1984} from the discussion of the quasiparticle.
The hierarchical state with filling factor $(m - 1/n)^{-1}$ ($n \in \mathbb{Z}$) is the $1/n$ Laughlin state of quaiparticles  
formed over the original $1/m$ one, where the ratio of the original fermion density to the quasiparticle density is $n$ to $1$.
Therefore we should tune the constraint imposing over the $1/m$ Laughlin state to
\begin{equation}
2\pi (Q_{j+1} - Q_{j}) = \frac{1}{n} (\Theta_{j+1} + \Theta_{j}).
\end{equation}
If we translate $(Q_{j+1} - Q_{j})$ to $\theta$ and $\varphi$ by using the original $1/m$ Laughlin state constraint, the relation between a gauge and a chiral transformation is
\begin{equation}
\Theta_{j+1} + \Theta_{j} = \frac{1}{m- \frac{1}{n}} \left(\Phi_{j+1} - \Phi_{j} \right) .
\end{equation}
Therefore we can find this state has the filling factor $\nu = (m - 1/n)^{-1}$.
From all these results, the bulk and edge effective theory, the fractional statistics of the quasiparticles, and the hierarchical structure, we can identify the constructed state as the $\nu = 1/m$ Laughlin state.

\section{Chiral spin liquid state in Coupled wire construction}
We can also construct the (Abelian) chiral spin liquid state \cite{KalmeyerLaughlin1987,Wen1989}, 
which is another example of topologically ordered states with anyonic excitations.  
The chiral spin liquid is considered as a $\nu=1/2$ fractional quantum Hall state of {\em bosonic} magnons (i.e., 
spin flip excitations). 
The bulk is effectively described by the level-2 U(1) Chern-Simons gauge theory 
and the quasiparticles are spinon excitations obeying semionic statistics \cite{KalmeyerLaughlin1987}.  
On the other hand, the edge state is the chiral Luttinger liquid for spinons. 

\subsection{Fictitious magnetic field through spin-orbit interaction}
Let us again begin with the theory for a single wire. 
We follow similar steps except that, due to the additional spin degrees of freedom, 
we need to introduce, instead of the spinless fermions, spin-1/2 (i.e., spin unpolarized) Dirac fermions
\begin{equation}
\psi_j = 
\begin{pmatrix}
\mathbf{R}_j \\
\mathbf{L}_j
\end{pmatrix}
=
\begin{pmatrix}
R_{+,j} \\
R_{-,j} \\
L_{+,j} \\
L_{-,j} \\
\end{pmatrix}
\; ,
\label{eqn:Dirac-CSL}
\end{equation}
where the $\pm$ represent the spin up and down.  

In the chiral spin liquids, in their original forms, time-reversal symmetry is broken as a consequence 
of many-body correlations.  Here we mimic the spin-spin interactions by the spin-selective ``gauge potential'' 
introduced by the spin-orbit interaction \cite{Kane-M-spin-Hall-05,Meng2015} 
\begin{equation}
\mathcal{H}_{\text{so},j} = -i  \lambda_{\text{SO}} \int\! dx \,  \psi^{\dagger}_{j}\sigma_{3} \partial_{x}\psi_{j} \; .
\label{eqn:spin-orbit}
\end{equation}
In the low-energy limit, this amounts to shifting the spatial derivative as (with $m$ being the mass of 
the spinful electron)
\begin{equation}
\partial_{x} \to \partial_{x} + i m \lambda_{\text{so}} \sigma_{3}   \; .
\end{equation}
We also add the Zeeman coupling that changes the chemical potentials in an opposite way for $+$ and $-$:
\begin{equation}
\begin{split}
\mathcal{H}_{\text{Z},j} &= -g \mu_{\text{B}} H  \int\! dx \, 
\psi_{j}^{\dagger} (\textbf{1} \otimes \sigma_{3}) \psi_{j}  \\
&\to -g \mu_{\text{B}} H \int\! dx \left(
R^{\dagger}_{j}\sigma_{3}R_{j} + L^{\dagger}_{j}\sigma_{3}L_{j} \right) \; .
\end{split}
\label{eqn:Zeeman}
\end{equation}

Summing up all these terms, we obtain the continuum-limit action for the $j$-th wire:
\begin{equation}
S_j = \int d^2x \bar{\psi}_j i \gamma^{\mu} \left( \partial_{\mu} - ik^{\text{F}}_{\mu} \gamma^\text{c}_5 
- ik^{\text{so},j}_{\mu} \gamma^\text{s}_5 - i k^{\text{Z},j}_{\mu} \gamma^\text{c}_5 \gamma^\text{s}_5\right) \psi_j  \; , 
\label{eqn:CSL-single-wire}
\end{equation}
where the Dirac conjugate is written as 
\begin{equation}
\bar{\psi}_j = \psi_j^{\dagger} \gamma^0 
\end{equation}
with the 4$\times$4 Dirac gamma matrices defined as
\begin{equation}
\gamma^0 = \sigma_{1}\otimes \textbf{1} = 
\begin{pmatrix}
0 & \textbf{1} \\
\textbf{1} & 0
\end{pmatrix}
,~~~~~ \gamma^1 = -i  \sigma_{2}\otimes \textbf{1} = 
\begin{pmatrix}
0 & -\textbf{1} \\
\textbf{1} & 0
\end{pmatrix}
\end{equation}
($\textbf{1}$ denotes the two-dimensional identity matrix).  
There are three ``external fields'' in the single-wire action \eqref{eqn:Dirac-CSL}:  
the first one expresses the Fermi momentum measured from the reference $k_{x}=-bj$:   
$k^{\text{F}}_{\mu} =(0, k_{\text{F}})$, which couples to 
\begin{equation}
\gamma^{\text{c}}_{5} =\gamma^0 \gamma^1 = \sigma_{3}\otimes \textbf{1} =
\begin{pmatrix}
\textbf{1} & 0 \\
0 & -\textbf{1}
\end{pmatrix}.
\end{equation}
As will be seen in Sec.~\ref{sec:Mott-limit}, the value of $k_{\text{F}}$ is determined by 
requiring that we can correctly reproduce the $S=1/2$ model in the Mott insulating limit.  

The second one $k^{\text{so},j}_{\mu} = (0, k^{j}_{\text{so}})\equiv(0,-m \lambda_{\text{so}})$ 
comes from the spin-orbit coupling \eqref{eqn:spin-orbit} 
which acts on the spin indices ($+$/$-$) through 
\begin{equation}
\gamma^{\text{s}}_{5} = \textbf{1}\otimes \sigma_{3} = 
\begin{pmatrix}
\sigma_3 & 0 \\
0 & \sigma_3
\end{pmatrix}.
\end{equation}
Note that the matrix $\gamma^{\text{s}}_5$ commutes with the other gamma matrices:
\begin{equation}
\left[ \gamma^{\mu}, \gamma^{\text{s}}_5\right] = 0 \; , \quad 
\left[ \gamma^{\text{c}}_5, \gamma^{\text{s}}_5\right] = 0.
\end{equation}

Last, the $k^{\text{Z},j}_{\mu}$ for the Zeeman coupling \eqref{eqn:Zeeman} is given by 
$k^{\text{Z},j}_{\mu}=(0,-g \mu_{\text{B}}H)$.  
After the gauge transformation similar to that introduced in Sec.~\ref{sec:bulk-FT-Laughlin} 
[see Eq.~]

Let us consider the meaning of these gamma matrices $\gamma^{\text{c}}_5$ and $\gamma^{\text{s}}_5$.  
To this end, it is convenient to apply the usual Abelian bosonization \cite{GiamarchText} to the single-wire action 
\eqref{eqn:CSL-single-wire}:
\begin{equation}
\begin{split}
& R_{a,j} = \frac{\kappa_{a,j}}{\sqrt{2\pi a_0}}
\exp\left(i  \phi_{a,j,\text{R}}\right)  
\\
& L_{a,j} = \frac{\kappa_{a,j}}{\sqrt{2\pi a_0}}
\exp\left(i \phi_{a,j,\text{L}}\right) \quad 
(a=+/-) \; ,
\end{split}
\label{eqn:CSL-boson2fermion}
\end{equation}
where $\kappa_{a,j}$ denotes the Klein factors that ensure the anti-commutation between 
the fermions with different spins ($a$) and $a_{0}$ is the cut-off parameter.  
Following the standard steps in Abelian bosonization, the low-energy physics of a single chain ($j$) 
is described by $\Phi_{a,j}=(\phi_{a,j,\text{R}}+\phi_{a,j,\text{L}})/2$, 
$\Theta_{a,j}=(\phi_{a,j,\text{R}}-\phi_{a,j,\text{L}})/2$. 

A gauge transformation $\psi_{j} \to \exp(i \varphi_{\text{c}})\psi_j$ of the Dirac fermion 
is related to the Josephson current $\partial_x \varphi_\text{c}$, while the chiral transformation 
$\psi_{j} \to \exp (i\theta_{\text{c}} \gamma^{\text{c}}_5)\psi_j$ is related 
to the charge-density $\rho_{\text{c}} = \partial_x \theta_{\text{c}} / \pi$ \citep{GiamarchText}.
Because of these features, we put the subscript ``c'' to the corresponding gamma matrix to 
represent the charge sector.
We can understand the $\gamma^s_5$ by the same manner.

A spin-dependent transformation, which we call a spinon gauge transformation,
\begin{equation}
\psi \rightarrow e^{i\varphi_\text{s} \gamma^\text{s}_5} \psi ,~~~~~ 
\bar{\psi} \rightarrow \bar{\psi} e^{-i\varphi_\text{s} \gamma^\text{s}_5}
\end{equation}
is related to the spin current.
The third transformation, dubbed a spinon chiral transformation,
\begin{equation}
\psi \rightarrow e^{i\theta_\text{s} \gamma^\text{c}_5 \gamma^\text{s}_5} \psi ,~~~~~ \bar{\psi} \rightarrow \bar{\psi} e^{i\theta_\text{s} \gamma^\text{c}_5 \gamma^\text{s}_5}
\end{equation}
is related to the spin density.
This is quite natural as the Zeeman coupling plays the role of ``chemical 
potential'' for the spin density 
\begin{equation}
\left( R^{\dagger}_{+}R_{+}-R^{\dagger}_{-}R_{-} \right) 
+ \left( L^{\dagger}_{+}L_{+}-L^{\dagger}_{-}L_{-}   \right)  \;.
\end{equation}
\subsection{Spin model in Mott-insulating limit}
\label{sec:Mott-limit}
\begin{figure}
\includegraphics[width=80mm]{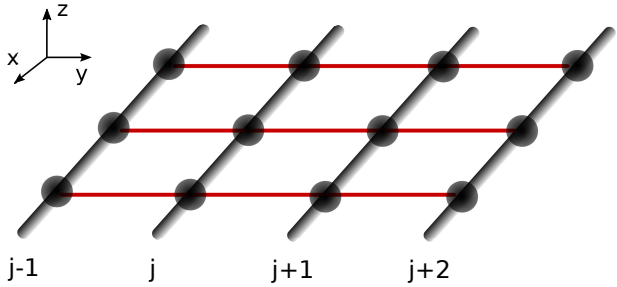} 
\caption{(Color online)
Mott-insulating limit.
The spin degrees of freeedom can exist only at the black spot and interact with each others through the Wilson lines.
\label{fig:Mott-state}}
\end{figure}

Let us prepare chains (running in the $x$-direction) with the lattice constant $l_x$  
and place them on a plane with the inter-wire distance $l_y$. 
In order to construct a spin model, we have to eliminate the charge fluctuations by stabilizing 
$4k_{\text{F}}$ charge density wave with strong electron-electron repulsion.   
As each site is occupied by exactly one electron in the Mott-insulating state (see Fig.~\ref{fig:Mott-state}), 
the average density of the Dirac fermion on a one dimensional wire $2k_\text{F} / \pi$ should be equal to $1/l_x$.
Therefore if we set $k_\text{F} = \pi / 2 l_x$ and tune the density fluctuation 
$\rho_\text{c} = \partial_x \theta_\text{c} / \pi$ to zero, a two-dimensional 
spin model whose lattice constants are $l_x$ and $l_y$ is obtained.
(we assume that $\theta_{\text{c}}$ is pinned to the appropriate value. )
The resulting action for the spinon degrees of freedom reads as
\begin{equation}
S_{j,\text{spinon}} = \int d^2x \bar{\psi '}_j i \gamma^{\mu}
\left( \partial_{\mu} - ik^{j,\text{so}}_{\mu} \gamma^\text{s}_5 - ik^{j, \text{Z}}_{\mu} \gamma^\text{c}_5 \gamma^\text{s}_5\right) \psi '_j .
\end{equation}
Note that $\psi^{\prime}_j$ and $\bar{\psi^{\prime}}_{j}$ are not the Dirac fermions.
For more precise treatment, we need to bosonize the original (spinful) Dirac fermions and move from the fermion basis to the bosonic basis (Luttinger liquids for charge and spin degrees of freedom).
If the charge sector is gapped out, then we can get the bosonic theory which contains only 
the spin degrees of freedom. 
Therefore the objects $\psi_j$ and $\bar{\psi}_{j}$ should be read as 
\begin{align}
&\psi '_j \sim e^{i \varphi_{\text{c}}} e^{i\varphi_\text{s} \gamma^\text{s}_5} e^{i\theta_\text{s} \gamma^\text{c}_5 \gamma^\text{s}_5} \kappa_j \\
&\bar{\psi '}_j \sim \bar{\kappa}_j e^{-i \varphi_{\text{c}}} e^{-i\varphi_\text{s} \gamma^\text{s}_5} e^{i\theta_\text{s} \gamma^\text{c}_5 \gamma^\text{s}_5},
\end{align}
where $\kappa_j$ and $\bar{\kappa}_j$ are the Klein factors.
However, all we need here is the form of the covariant derivative and the transformation 
properties of the objects $\psi '_j$ and $\bar{\psi '}_{j}$.

Next let us introduce a Wilson line and an interaction between two wires.
The form of the Wilson line is
\begin{equation}
\bar{\psi '}_{j+1} \exp \left[ i\int^{(j+1)l_y}_{j l_y} dy \left( k^\text{so}_y \gamma^\text{s}_5 
+ a^{\text{s}}_y \gamma^\text{c}_5 \gamma^\text{s}_5 \right) \right]\psi '_j ,
\end{equation}
where $k^{\text{so}}_y$ is the spin-orbit coupling on the plane and $a^{\text{s}}_y$ is an unknown field.
Under a spinon gauge transformation, $k^{\text{so}}_y$ behaves like a gauge field.  
As in Sec.~\ref{sec:CWC-for-Laughlin}, we assume that we have tailored the inter-chain interactions in such a way that 
the spin-fields are constrained as
\begin{equation}
\theta^{j+1}_\text{s} + \theta^{j}_\text{s} = \frac{1}{2} \left(\varphi^{j+1}_\text{s} - \varphi^{j}_\text{s} \right) . 
\label{eqn:constraint-CSL}
\end{equation}
\subsection{Effective theories and quasiparticles}
As in the construction of the Laughlin state in Sec.~\ref{sec:bulk-FT-Laughlin}, 
if there is a constraint of this type [see Eq.~\eqref{eqn:constraint-CSL}], 
$a_y$ behaves like a spinon gauge field under a spinon chiral transformation.
Therefore, it is possible to examine the role of $a^{\text{s}}_y$ in a similar way to the Laughlin case.
The Zeeman coupling is the average density of spin and $\partial_x a^{\text{s}}_y$ is the spin density fluctuation.
Therefore $a^{\text{s}}_y$ is the U(1) Chern-Simons gauge field and the bulk effective theory is the level-2 U(1) Chern-Simons gauge theory.
The gapless excitation in the bulk is suppressed by the interaction and the gapped excitation is a spinon, 
which has fractional statistics.
On the other hand, the edge state is described by the chiral Luttinger liquid.

\vspace{5mm}

\section{Conclusion}
In this paper, we have developed the reformulation of CWC by using the Wilson line for the local gauge invariance.
We have taken the Laughlin state and the chiral spin liquid state for the examples and derived the Chern-Simons gauge theories in the bulk for each.
The quasiparticles and their statistics have been naturally derived in the bulk space.
For these reasons, we conclude that the array of one-dimensional system under an appropreate non-local interaction can be identified with the two-dimensional topologically ordered state.

Finally, we would like to refer to a future problem about the CWC.
It is not trivial how the entanglement structure of the whole system is constructed by the inter-wire interaction.
The entanglement entropy of the Tomonaga-Luttinger liquid defined on the each wire has the form $S_{\text{TL}}=(1/3)\log (l/\varepsilon)$.
On the other hand, that of the level-m U(1) Chern-Simons gauge theory has the topological entanglement entropy term $ - \log \sqrt{m}$.
The difference between them should be understood in the inter-wire interaction point of view.
However, because the CWC formulated above is analyzed in the semi-classical limit, now we are not able to calculate quantum measurements like the entanglement entropy.

\begin{acknowledgments}
We thank Kazuma Nagao for helpful comments on this work.
On of the authors (K.T.) was supported in part by JSPS KAKENHI Grant No.~24540402 and No.~15K05211.  
\end{acknowledgments}

\bibliographystyle{apsrev4-1}
\bibliography{CWC_bibliography.bib}

\end{document}